# The Effect of mechanical alloying on Electrical properties of BaTiO3 Nano Crystals

Masoud Mollaee[1,2,*], Mahmoud R. Roknabadi[1], Mahdieh Zaboli[1]

[1]Department of Physics, Ferdowsi University of Mashhad, Iran

Department of Physics, University of Texas at El Paso, USA

[*]mmollaee@miners.utep.edu

## Abstract

In this paper, electrical properties of BaTiO3 nano crystals have been studied, Barium Titanate Nano Crystals are made by mechanical alloying method in a ball mill of SPEX 8000. In order to find Curie temperature, dielectric constant variation via temperature and hysteresis loops are investigated. Our results show that, there is a relationship between the time of milling and the curie temperature .this means with increasing milling time, grain size will decrease, consequently the curie temperature will decrease. In addition, with increasing temperature up to the Curie temperature, the hysteresis loop of samples decrease and at the Curie temperature the hysteresis loop changes to a straight line.

## 1. Introduction

Ferroelectric materials is one of the research interesting subjects of scientists who have been working since 1920 [1] and obtaining new achievements in different aspects of technology. Ferroelectric materials have been made a close study of electrical , electromagnetic ,nonlinear optical properties[2-3]. These days, they have been under investigation due to the prospect that the stable polarization states could be used to encode the 1 and 0 of the Boolean algebra that form the basis of memory and logic circuitry (FRAM). The family of complex ferroelectric oxides such as BaTiO3, Pb(Zr,Ti)O3 has far reaching applications in the electronics industry for transducers, actuators, and high-k dielectrics,[4-9].There has been growing trend toward preparing nano ferroelectric materials and investigation into the nano properties as well as thin films[4,7,10,11] . Wul and Goldman discovered the ferroelectricity in Barium Titanate in 1945[12].it has been one of the most exhaustively studied materials as well it is one of the materials most widely used in modern electronic and technical devices[11,13-15] . Barium Taitanate is used as a ferroelectric material with a high dielectric constant, it has been widely studied for application in capacitor, varistor and even random access memory (RAM) with developing ultra large scaled integrated circuits (ULSI) and solar cells [16-18], ceramics of Barium Taitanate are also used for making multilayer capacitors, transformators and Temperature sensors[19-21]. In a ferroelectric material there is a specific temperature which at the upper of that the spontaneous Polarization is vanished. In fact at this temperature the crystal structure of material changes from ferroelectricity to parraelectricity. This critical temperature is called Curie temperature. Every ferroelectric material has its specific curie temperature. The curie temperature of a material possesses some data about the structure and characterization of sample such as grain size, density, purity [22-23].The Curie temperature is determined by diagrams of dielectric constants variation via temperature and hysteresis loop. Increase in temperature contributes to the increase in the dielectric constant up to the curie temperature.

The study of size effect in ferroelectric systems has lately very important because of their potential applications. It was found that the dielectric properties of BaTiO3 strongly dependent of its grain size. Coarse-grained ceramics (20–50μm) of pure Barium Titanate showed lower dielectric constant at room temperature then fine grained (0.8–1μm). Many authors considered that when the grain size is lower than 700 nm, the lattice of Barium Titanate changes from tetragonal to pseudocubic, and the dielectric constant value is very low, in the doped BT that effect is more complex [24-27].

In the present study, we have investigated the curie temperature changes as result of different milling and different grained size of Barium Titanate nano crystals.

## 2. Experimental Procedure

By BaTiO3 nano powders which are made by mechanical alloying in different times of 3, 6, 10 and 12, for tablets are produced by hydraulic press apparatus .then the tablets are sintered for two hours at 1200 C.to measuring at the first the surface of samples are polished then the electrodes are connected to the tablets. To make a better connection samples are put in an oven for half an hour in 100 C. The Vetson Bridge is used to measure samples dielectric constants in terms of temperature.

In figure 1 points A and B are connected to the oscilloscope, if the total impedance of the circuit obey the following equation.

$$Z_1 R_1 = Z_2 R_2 \qquad (1)$$

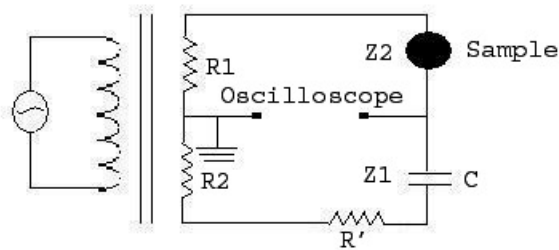

Figure1.Veston Bridge used in measuring of dielectric constant.

Then the potential difference between A and B will be zero. The resistance R' and the capacitor C' are variable and we can adjust them where as the equation 1 is established. In this condition, sample impedance will be equal with impedance of the branch including C' and R':

$$Z' = R' - \frac{j}{C'W} \rightarrow Z' = Z \qquad (2)$$

To get sample impedance Z, consider the sample as capacitor C and resistance R which are connected parallel where:

$$Z = \frac{R - jR^2CW}{1 + R^2C^2W^2} \quad (3)$$

By definition, dielectric dissipation is equal to ratio of current transition of resistance to current transition of capacitor where:

$$\tan\alpha = \frac{I_R}{I_C} = \frac{1}{RCW} \quad (4)$$

Regarding to equations 2 and 4 and because of sample capacity is about pico farad we have:

$$C = C' \quad (5)$$

$$R = \frac{1}{R'C'W} \quad (6)$$

So, the dielectric constant of the material is

$$\epsilon = \frac{Cd}{A} \quad (7)$$

The hysteresis loop for ferroelectric material is dependent on the temperature and frequency.

It is easy to prove that the electrical field can be calculated by the following formula from figure2:

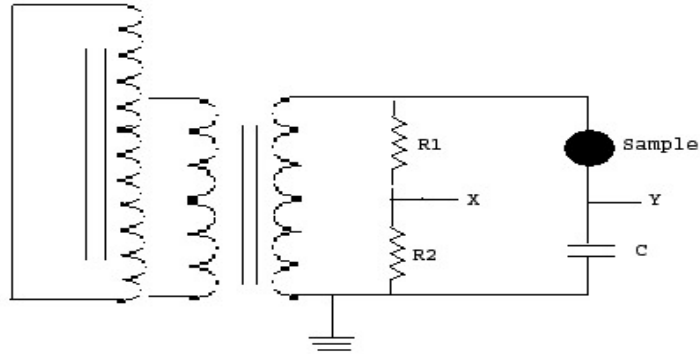

$$E = \frac{|R_1 - R_2|}{R_2}\frac{V_x}{d} \quad (8)$$

For measuring polarization, the capacitor $C_1$ is connected to the sample as a series circuit. If $C_1$ is greater than the C, then the applied voltage in this branch will appear between the sample. Electrodes and $C_1$ have no effect on the electric field determination because we have:

$$Z = \frac{1}{jCW} \quad (9)$$

$$Z_{C_1} = \frac{1}{jC_1W} \quad (10)$$

If $C_1 \gg C$, then the impedance of the $C_1$ capacitor will be less than the sample impedance. In this case the whole voltage will appear at the end of samples. The polarization per unit volume is:

$$P = \frac{Q}{A} \quad (11)$$

Where Q is the collected charge on the sample surface, and A is the area of the sample surface. Because the sample capacitor C and $C_1$ are connected as a series, the charges on them are:

$$Q = C_1 V_Y \qquad (12)$$

If we put the oscilloscope in the x-y position, the hysteresis loop will be appear. By using equation 8, 11,12 we can calculate the polarization and electric field, and then the spontaneous polarization $P_S$, retains polarization $P_R$ and the resonance field using the hysteresis loop will be calculated.

## 3. Results and Discussion

The Dielectric properties of BaTiO3 ceramic, which has been produced by mechanical alloying, have been studied using the circuits of figure 1 and 2. At first, the sample was put in a vessel of silicon oil which was on an electrical heater and stirred well by a magnetic force.

Using circuit 1, the diagrams of dielectric constant versus temperature (20-100 C) for different milling time were plotted.

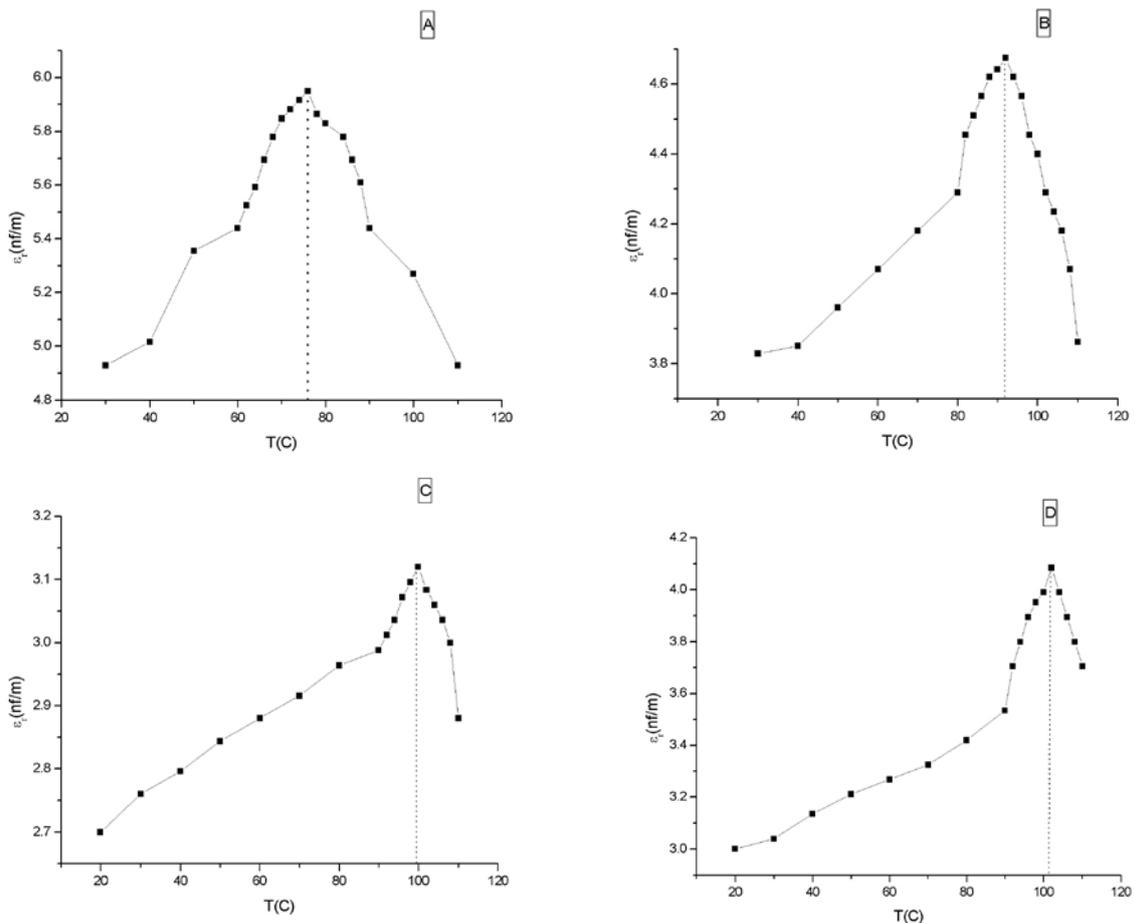

Figure3. the variation of dielectric constant versus temperature for different milling time. (a) 12 hours, (b) 10 hours, (c) 6 hours and (d) 3 hours.

As it can be expected, plots of dielectric constant versus temperature show that in transition, temperature ($T_c$) from ferroelectricity to parraelectricity, plots have their maximum values.

As it can be seen from figure 3 the milling processes has an important effect on the crystal structure, and the properties of BaTiO3. As it is seen there is a relation between milling time and Curie temperature. With increasing milling time the Curie temperature goes to the lower value (for example the sample with 3 hours milling time $T_C = 102C$ and for the sample with 12 hours milling time is equal to $T_C = 76C$)

We have also used the circuit of figure 2 to produce hysteresis loops for the tablet samples in 1 kHz frequency.

As we can investigate from figures 4 to 7, with increasing temperature the slope of plots increase and their width decrease. Exactly at the Curie temperature the hysteresis loops change to a straight line.

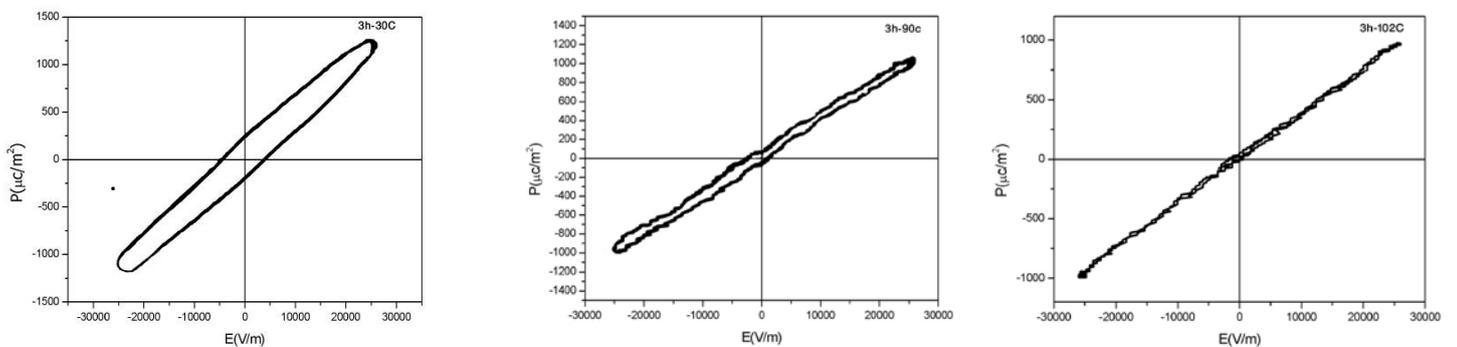

Figure 4- hysterias loop for BaTiO3 table made from powder with 3 hours alloying and at 1200C sintering 30C,90C,102C respectively

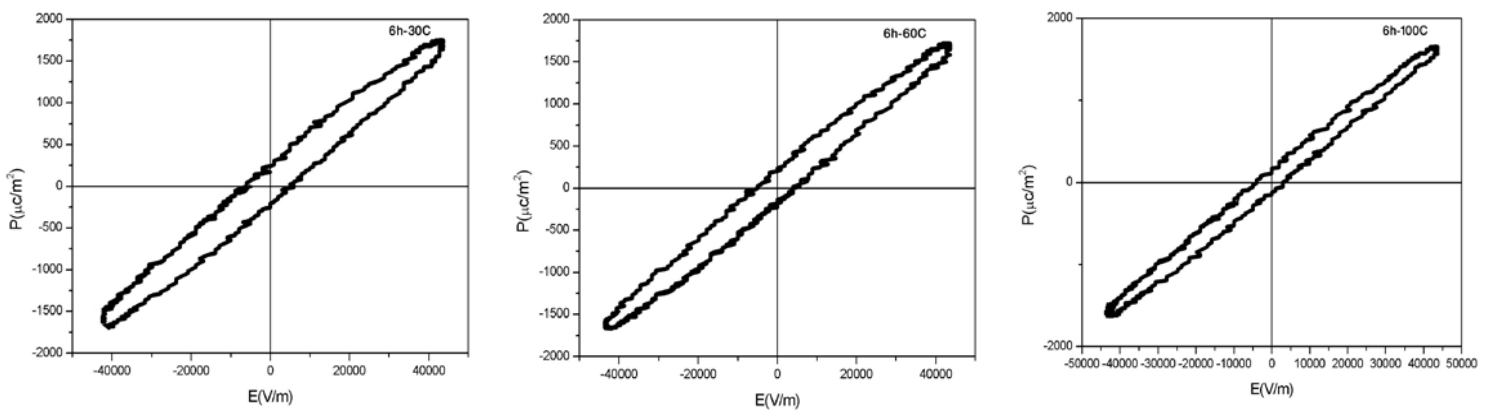

Figure5- hysteresis loop for BaTiO3 tablet made from powder with 6 hours alloying and at 1200C sintering.

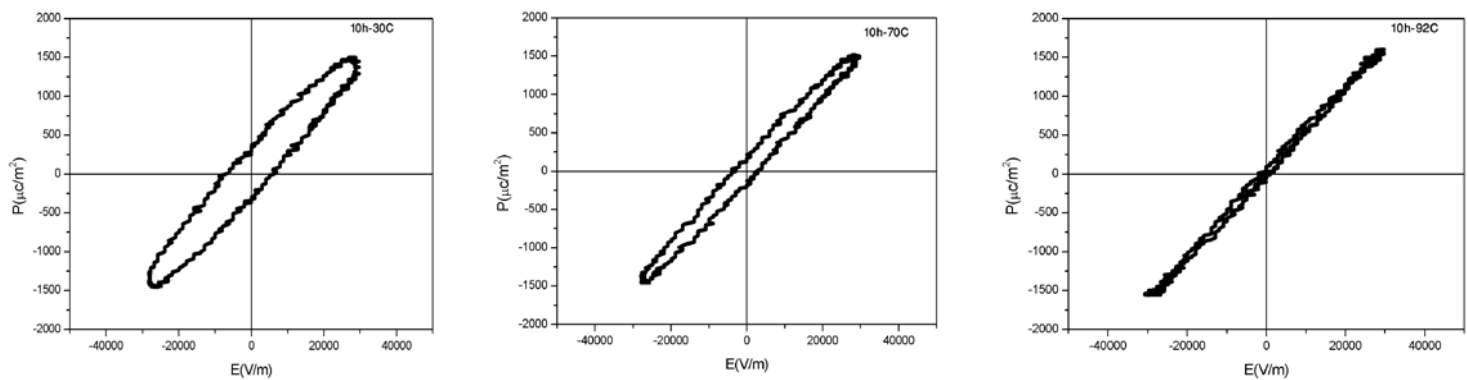

Figure6- hysterias loop for BaTiO3 tablet made from powder with 10 hours allying and at 1200c sintering.

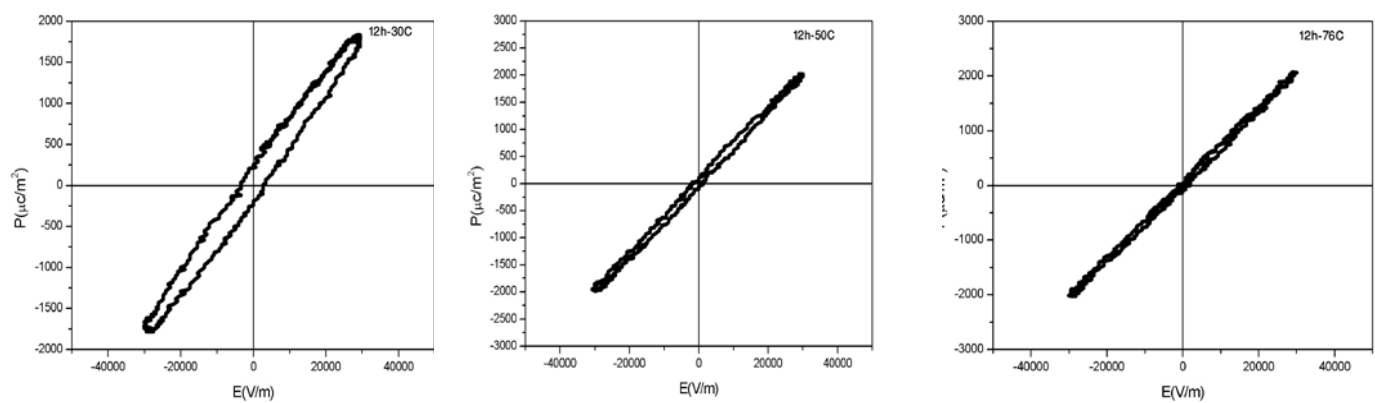

Figure7- hysterias loop for BaTiO3 tablet made from powder with 12 hours alloying and at 1200C sintering.

This phenomenon is due to decreasing of particle size. Because with decreasing of particle size, the number of dipoles of every domain is also decreased .even it is possible to have a domain only with one dipole. In this case, the dipoles need a lower electric field to be parallel with the external field which results in the Curie temperature also being decreased.

## Summary

In this study, we have prepared the tablets of BaTiO3 powder which was made by mechanical alloying during the time of 3,6,10 and 12 hours, at 1200C sintering temperature. The variation of dielectric constant versus temperature and the hysteresis loop of the samples are measured at different times.
Our results show that with increasing milling time, Curie temperature is decreased. It also shows that the width of hysteresis loops samples are decreased as the temperature is increased and it reaches to zero at Curie temperature.

## References

[1] J.Valasek , Physical Review. 15 (1920) 537.
[2] J. Yeon, P. Shiv Halasyamani, I.V. Kityk, Materials Letter. 62 (2007) 1082.

[3] P.K. Roy, J. Bera, Materials Chemistry and Physics. 132 (2012) 354.
[4] Wan Soo Yun, Jeffrey J. Urban, Qian Gu, and Hongkun Park, Nano letters. 2 (2002) 447.
[5] Yuanqing Cui, Zheng Zhong , Mechanics of Materials. 45 (2012) 61.
[6] M.C. Kao, H.Z. Chen, S.L. Young, B.N. Chuang, W.W. Jiang, J.S. Song, S.S. Jhan, J.L. Chiang, L.T. Wu, Journal of crystal growth. 338 (2012) 139.
[7] Stephen O'Brien, Louis Brus, and Christopher B. Murray, J. Am. Chem. Soc. 123 (2001) 12085.
[8] Wul,W.Goldman,I.M.C.R,  Acad.Sci.Russ. 40 (1946) 139.
[9] Ali Hussain RESHAK, Sushil AULUCK, Ivan KITYK, Japanese Journal of Applied Physics. 47 (2008) 5516.
[10] M. Piasecki,I.V. Kityk, P. Bragiel,K. Ozga,V. Kapustiany,B. Sahraoui.Chemical Physics Letters. 433 (2006) 136.
[11] M.I.Kolinko, I.V.Kityk, R.Y.Bibikov & J.Kasperczyk, Journ. Materials Science Letters. 15 (1996) 803.
[12] Hill, N. A. J. Phys. Chem. B. 104 (2000) 6694.
[13] C. Pitham, D. Hennings, and R. Wases, Int. J. Appl.Ceram. Technol. 2 (2005) 1.
[14] M. Dawber, K. M. Rabe, and J. F. Scott, Rev. Mod.Phys. 77 (2005) 1083.
[15] Michael Z.-C. Hu, Vino Kurian , E. Andrew Payzant, Claudia J. Rawn,Rodney D. Hunt,Powder Technology. 110 (2000) 2.
[16] Chen-Feng Kao and Chiao-Ling Yang, Journal of the European Ceramic Society. 19 (1999) 1365.
[17] Yoshimura, M., Yoo Seung-Eul ,Hayashi  and Nobuo Ishizawa. Japanese J. Applied Physics. 28 (1989) 2007.
[18] Min Zhong, Jingying Shi, Wenhua Zhang, Hongxian Han, Can Li, Materials Science and Engineering B. 176 (2011) 1115.
[19] Zhibin Wu, Masahiro Yoshimura, Solid State Ionics. 122 (1999) 161.
[20] Giuseppe Vasta, Timothy J. Jackson, Edward Tarte, Thin solids film. 520 (2012) 4506.
[21] Jinseong Kim, Dowan Kim, Taimin Noh, Byungmin Ahn, Heesoo Lee, Materials Science and Engineering B. 176 (2011) 1227.
[22] Y. Su, G.J. Weng, Journal of the Mechanics and Physics of Solids. 53 (2005) 2071.
[23] Yue-Ming Li, Liang Cheng, Xing-Yong Gu, Yu-Ping Zhang, Run-Hua Liao, Journal of Materials Processing Technology. 197 (2008) 170.
[24] B.D. Stojanovic,C.R. Foschini, M.A. Zaghete, F.O.S. Veira,K.A. Peron, M. Cilense, J.A. Varela, Journal of Materials Processing Technology. 143 (2003) 802.
[25]B. Jaffe, W.R. Cook, Piezoelectric Ceramics, Academic Press,London, New York, 1971.
[26] G.H. Heartling, J. Am. Ceram. Soc. 82 (1999) 797.
[27] G. Bush, Ferroelectrics. 4 (1987) 267.